\documentclass[amsmath,amssymb,twocolumn]{revtex4}
\usepackage{graphicx}
\usepackage{amscd,amsmath,amsthm,amsfonts,amssymb}

\def\sech{{\rm sech}}

\def\ri{{\rm i}}

\def\re{{\rm e}}
\def\PT{$\mathcal{PT}$}

\begin{document}
\title{All-real spectra in optical systems with arbitrary gain and loss distributions}
\author{Sean Nixon and Jianke Yang}

\address{Department of Mathematics and Statistics, University of Vermont, Burlington, VT 05401, USA}

\begin{abstract}
A method for constructing optical potentials with an arbitrary distribution of gain and loss and completely real spectrum is presented. For each arbitrary distribution of gain and loss, several classes of refractive-index profiles with freely tunable parameters are obtained such that the resulting complex potentials, although being non-parity-time-symmetric in general, still feature all-real spectra for a wide range of tuning parameters. When these refractive indices are tuned below certain thresholds, phase transition can occur, where complex-conjugate pairs of eigenvalues appear in the spectrum. These non-parity-time-symmetric complex potentials generalize the concept of parity-time-symmetric potentials to allow for more flexible gain and loss distributions while still maintaining all-real spectra and the phenomenon of phase transition.
\end{abstract}


\maketitle

Parity-time (\PT) symmetric optics is a frontier of current research. \PT symmetry, first introduced by Bender and Boettcher \cite{Bender1998} as a non-Hermitian generalization of quantum mechanics, involves the study of potentials which, though complex, still possess an all-real spectrum. This concept later spread to optics, where an even refractive index profile together with an odd gain-loss landscape constitutes a \PT-symmetric system and could feature all-real spectrum \cite{Musslimani2008}. In this optical context, \PT symmetry was demonstrated experimentally for the first time \cite{Ruter_2010,Regensburger_2012}. A distinctive phenomenon in \PT-symmetric systems is phase transition, where the spectrum changes from all-real to partially complex when the gain-loss component (relative to the index of refraction) exceeds a certain threshold \cite{Bender1998,Musslimani2008,Ruter_2010,Regensburger_2012,Ahmed2001}. This phase transition has been utilized for many emerging applications of \PT optics, such as single-mode \PT lasers \cite{PTlaser_Zhang,PTlaser_CREOL} and unidirectional reflectionless optical devices \cite{Feng2013}. When nonlinearity is introduced into \PT-symmetric systems, the interplay between \PT symmetry and nonlinearity yields additional interesting properties which are
being actively explored \cite{Driben2011, Zezyulin2012a, Nixon2012b, Segev2013, Wimmer2015}.
\PT-symmetric systems break the boundaries between traditional conservative and dissipative systems and open up new exciting research territories. In addition, \PT symmetry makes loss useful, which is quite enlightening.

Optical \PT symmetry requires the refractive index to be even and gain-loss profile to be odd in space, which is quite restrictive. This leads us to the natural question, how can we relax the \PT-symmetry requirement of the optical potential, while still preserving the all-real spectrum? Recently, several examples of non-\PT-symmetric complex potentials with real spectra have been reported \cite{Cannata1998,SUSY2013,Tsoy2014}. In \cite{Cannata1998,SUSY2013}, families of non-\PT potentials with real spectra were constructed by supersymmetry. In \cite{Tsoy2014}, a class of such non-\PT potentials were constructed by relating the Schr\"{o}dinger eigenvalue problem to the Zakharov-Shabat eigenvalue problem. In the latter case, the resulting non-\PT potentials not only admit all-real spectra, but also support continuous families of soliton modes \cite{Tsoy2014,Konotop_OL2014} and allow for symmetry breaking of soliton families \cite{Yang_OL2004}.

In this article, we outline a more general approach to finding non-\PT-symmetric complex potentials with real spectra by relating the Schr\"odinger operator with its complex conjugate. Using this approach, we derive several classes of complex potentials whose gain-loss distribution is completely arbitrary (even including cases with only gain and no loss), yet, surprisingly, their spectra can be all-real. These potentials contain those reported in \cite{Tsoy2014} as special cases, but they also contain new types of non-\PT potentials which cannot be obtained by the methods in \cite{Cannata1998,SUSY2013,Tsoy2014}. For an arbitrary gain-loss distribution, the refractive index of our complex potential contains free parameters which make it tunable. Tuning the refractive index below a certain threshold can result in the spectrum undergoing a phase transition, where the spectrum changes from all-real to partially-complex. These complex potentials with arbitrary gain-loss profiles and all-real spectra or phase transition provide a more flexible alternative to \PT-symmetric potentials for future applications.

Linear paraxial propagation of light in an optical waveguide is governed by the Schr\"{o}dinger equation
\begin{equation} \label{e:SE}
\ri \Psi_z + \Psi_{xx} + V(x) \Psi=0,
\end{equation}
where $z$ is the distance of propagation, $x$ is the transverse coordinate,
\begin{equation} \label{e:V}
V(x)=n(x) -i G(x)
\end{equation}
is a complex potential whose real part $n(x)$ is the index of refraction and whose imaginary part $G(x)$ represents gain and loss in the waveguide ($G>0$ represents gain and $G<0$ represents loss).  Looking for eigenmodes of the form $\Psi = \re^{\ri \mu z} \psi(x)$ we arrive at the eigenvalue problem
\begin{equation}
L\psi=\mu \psi,
\end{equation}
where
$L = \partial_{xx} + V(x)$ and $\mu$ is an eigenvalue. We will show that for an arbitrary choice of the gain-loss profile $G(x)$,
there exist wide families of refractive-index profiles $n(x)$ which result in a completely real spectrum.

Our approach is based on the following observation: if there exists an operator $\eta$ such that $L$ and its complex conjugate $L^*$ are related by a similarity relation
\begin{equation} \label{Eq:LSimilar}
\eta L =  L^* \eta,
\end{equation}
then the eigenvalues of $L$ come in conjugate pairs under certain mild conditions. To see this, take any eigenvalue and eigenfunction pair $(\mu, \psi)$ of the operator $L$, i.e., $L\psi=\mu \psi$. Obviously $\mu^*$ is an eigenvalue of $L^*$ since $L^* \psi^*=\mu^*\psi^*$. Then using the similarity relation (\ref{Eq:LSimilar}) we get
\[
L (\eta^* \psi^*)= (L\eta^*)\psi^*=(\eta^*L^*)\psi^*=\eta^*(\mu^*\psi^*)= \mu^* (\eta^* \psi^*),
\]
i.e., $\mu^*$ is also an eigenvalue of $L$ (as long as $\eta\psi\ne 0$). Thus, eigenvalues of $L$ come in conjugate pairs if the kernel of $\eta$ is empty (i.e., $\eta \psi =0$ implies $\psi=0$), which is a sufficient but not necessary condition.

For the Schr\"odinger operator $L$, $L^*$ is equal to its Hermitian
$L^\dagger$, thus the similarity condition (\ref{Eq:LSimilar}) is
the same as $\eta L =  L^\dagger \eta$. This condition resembles the
condition for pseudo-Hermiticity, which comes up in quantum
mechanics \cite{Mostaf2010}, the difference being a looser set of
conditions on $\eta$. For pseudo-Hermiticity $\eta$ must be both
invertible and Hermitian \cite{Mostaf2010}, however these extra
conditions are unnecessary for our purposes. As we will see later in
the text, $\eta$ in our consideration is generally non-Hermitian,
thus operators $L$ satisfying the similarity relation
(\ref{Eq:LSimilar}) may be non-pseudo-Hermitian.

The fact that eigenvalues of $L$ come in conjugate pairs means that simple real eigenvalues of $L$ cannot bifurcate off the real axis, and complex eigenvalues can only appear when pairs of simple real eigenvalues collide. Thus, if the spectrum of $L$ is completely real for a certain potential, then it will remain completely real under general deformations of the potential until real eigenvalues collide (where phase transition occurs). Thus, we expect that under the similarity condition (\ref{Eq:LSimilar}) and some weak conditions on the kernel of $\eta$, there will be large families of potentials which exhibit real spectra. In addition, phase transition is possible.

The above qualitative features of the spectrum closely resemble those of \PT-symmetric operators \cite{Bender1998,Musslimani2008,Ahmed2001}. The reason for this is that operators satisfying the similarity condition (\ref{Eq:LSimilar}) contain \PT-symmetric operators as special cases. Specifically, when we let $\eta = \mathcal{P}\equiv x \mapsto -x$ (the parity operator), then this $\eta$ operator has an empty kernel, and the similarity condition (\ref{Eq:LSimilar}) yields $V(-x)  =V^*(x)$, which recovers the well-known class of \PT-symmetric Schr\"odinger operators $L=\partial_{xx} + V(x)$.   For $\mathcal{PT}$-symmetric operators, the condition on the complex potential $V(x)$ is rather strict in that the gain-loss distribution must be an odd function. However, when branching out to different choices of $\eta$, a completely real spectrum can be obtained for an arbitrary choice of the gain-loss distribution by the judicious construction of the index of refraction. This will be demonstrated below where $\eta$ is chosen as a differential operator.

First, we consider the simplest choice of a differential $\eta$ operator,
\begin{equation} \label{e:eta1}
\eta = \partial_x + a(x).
\end{equation}
Substituting this $\eta$ and the potential (\ref{e:V}) into the
similarity condition (\ref{Eq:LSimilar}), we get the following two
equations
\begin{equation} \label{e:iG}
a_x=-iG, \qquad a_{xx}-V_x=(a^2)_x.
\end{equation}
The second equation can be integrated once, and we get
\[
a_{x}-V=a^2+\xi_0,
\]
where $\xi_0$ is a constant. Utilizing (\ref{e:iG}), this equation becomes
\begin{equation} \label{e:n}
n=-a^2-\xi_0.
\end{equation}
Since $G$ and $n$ are real functions, Eqs. (\ref{e:iG})-(\ref{e:n})
show that $a(x)$ is a purely imaginary function, and $\xi_0$ is a
real constant. Denoting $a(x)=ig(x)$, where $g(x)$ is an arbitrary
real function, we get $G=-g'(x)$ and $n=g^2(x)-\xi_0$. The constant
$\xi_0$ can be eliminated by a gauge transformation to Eq.
(\ref{e:SE}), thus the resulting complex potential is
\begin{equation} \label{e:V1}
V(x)=g^2(x)+ig'(x).
\end{equation}
This potential reproduces the one obtained in \cite{Tsoy2014} by a different method. For this potential, one can verify that the operator $L$ factors as $L = \eta^* \eta$.

It is important to notice that $g(x)$ in the above potential is an arbitrary function, meaning the gain-loss distribution $G(x)=-g'(x)$ is also arbitrary. Nonetheless, this potential satisfies the similarity relation (\ref{Eq:LSimilar}), thus its spectrum can be all-real. This all-real spectrum for the underlying non-\PT-symmetric potential is possible since its refractive index $n(x)$ is judiciously chosen when the gain-loss profile is given. Also notice that for each given gain-loss profile $G(x)$, the function $g(x)$ is not unique. Indeed, from $g'(x)=-G(x)$, we get
\begin{equation} \label{e:gxc0}
g(x)=\hat{g}(x)+c_0,
\end{equation}
where $\hat{g}(x)\equiv -\int_{-\infty}^x G(y)dy$ is the cumulative gain and loss, and $c_0$ is a free real parameter. Inserting this $g(x)$ expression into (\ref{e:V1}) and removing an overall constant $c_0^2$, this potential becomes
\begin{equation} \label{Eq:PotentialTypeI}
V(x)= \hat{g}^2(x) +2 \hspace{0.03cm} c_0 \hat{g}(x) - \ri G(x).
\end{equation}
It is seen that for any given $G(x)$, the refractive index of this potential has a free parameter $c_0$. As we will show below, tuning this $c_0$ parameter can change the spectrum from all-real to partially complex and thus induce phase transition.

\small
\begin{figure}[htb]
\includegraphics[width=8.3cm]{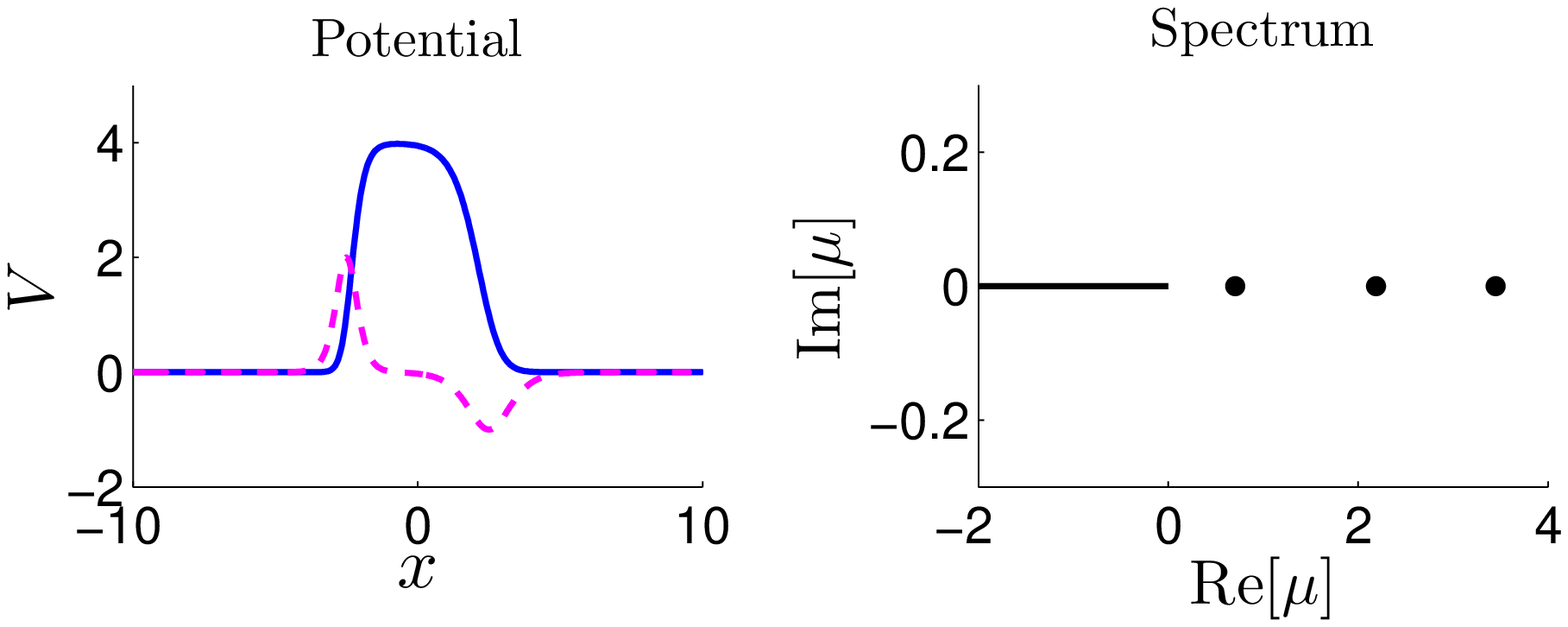}

\vspace{0.1cm}
\includegraphics[width=8.3cm]{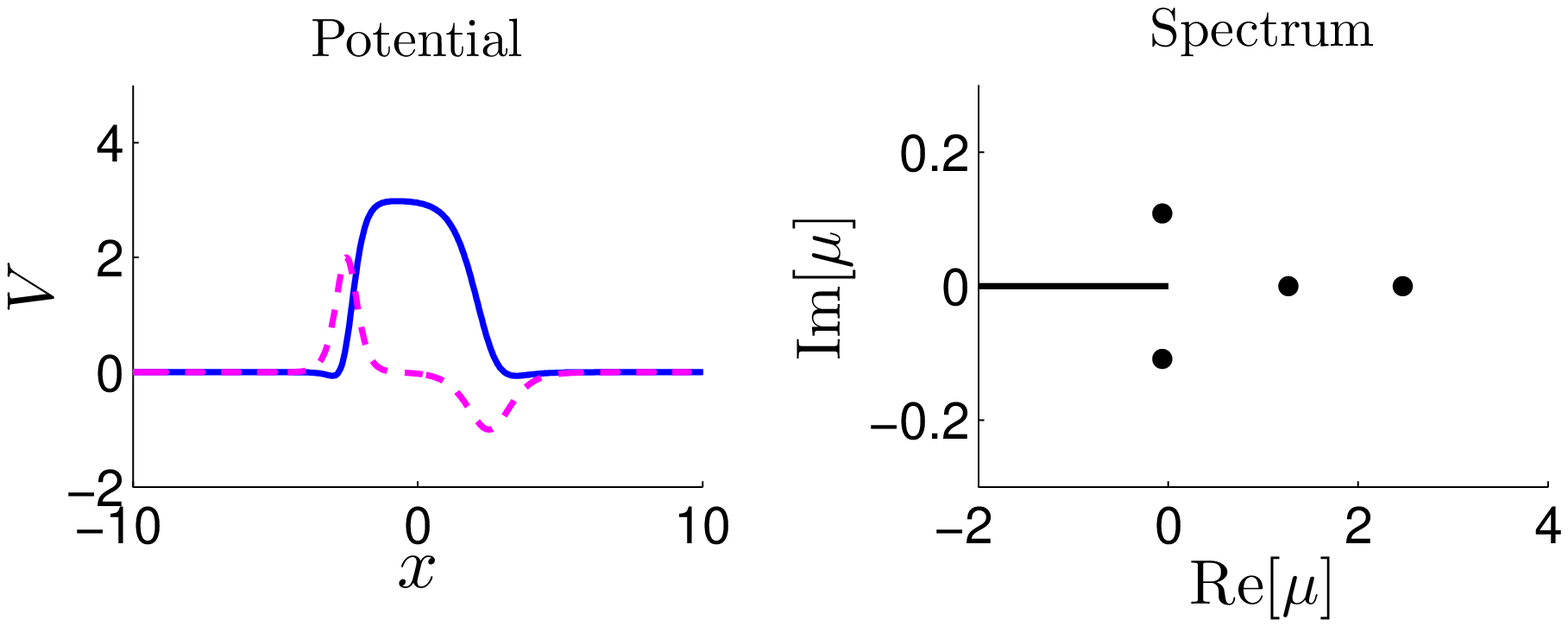}
\caption{Spectra of potentials (\ref{Eq:PotentialTypeI}) with gain-loss profile (\ref{Eq:GL}). Upper row: this potential with $c_0 = 0$; lower row: this potential with $c_0 = -0.3$. In the potentials, the solid blue line is Re($V$), i.e., the index of refraction $n(x)$, and the dashed pink line is Im($V$)$=-G(x)$, where $G(x)$ is the gain-loss profile. }
\label{Fig:FirstOrder}
\end{figure}
\normalsize

For an example, we take the gain-loss profile
\begin{equation} \label{Eq:GL}
G(x)  = -2\hspace{0.06cm} \sech^2 \hspace{-0.02cm} [2(x+2.5)] + \sech^2 \hspace{-0.02cm} (x-2.5),
\end{equation}
consisting a narrow region of strong loss followed by a wide region of weak gain.
This gain-loss profile is plotted in Fig. \ref{Fig:FirstOrder}. Obviously this profile is not an odd function, and therefore the resulting complex potential is not \PT-symmetric. But the total gain and loss are still balanced, i.e., $\int_{-\infty}^\infty G(x)dx=0$. In the corresponding potential (\ref{Eq:PotentialTypeI}), the constant $c_0$ tunes the magnitude of the refractive index (higher $c_0$ yields higher index). In Fig. \ref{Fig:FirstOrder}, we show two possible refractive indices of the form \eqref{Eq:PotentialTypeI} for the $c_0$ values of 0 and $-0.3$ respectively.  On top, the resulting potential with $c_0=0$ has completely real spectrum, and increasing $c_0$ will maintain the reality of the spectrum as more discrete eigenvalues bifurcate off the continuous spectrum. As $c_0$ is decreased, the spectrum will undergo a phase transition, resulting in a pair of eigenvalues bifurcating into the complex plane. This can be seen in the bottom of Fig. \ref{Fig:FirstOrder} with $c_0=-0.3$.

This first family of potentials (\ref{e:V1}) comes from taking the simplest form of a differential $\eta$ operator [i.e., a first-order operator (\ref{e:eta1})]. By increasing the order of this differential operator, more families of potentials arise. Let
\begin{equation}
\eta = \partial_{xx} + a(x) \hspace{0.04cm} \partial_{x} + b(x).
\end{equation}
Inserting this $\eta$ into the similarity condition (\ref{Eq:LSimilar}) and collecting coefficients of the same order of derivatives on the two sides of this condition, we get
\begin{center}
\begin{tabular}{|c||c|c|}
\hline
&$\eta L$ &$L^* \eta$ \\
\hline
$\partial_x^4$ &$ 1$ & $1$\\
$\partial_x^3$ & $a$ & $a$  \\
$\partial_x^2$ & $ V $ & $V^*+ 2a'$\\
$\partial_x^1$ &$V a+2V'$&$V^* a+ a''+ 2b'$\\
$\partial_x^0$ &$Vb+ V' a+V''$&$V^*b+ b''$\\
\hline
\end{tabular}
\end{center}
Setting these coefficients in $\eta L$ and $L^*\eta$ to match each other, we get a system of equations which can be solved from top to bottom. From the $\partial_x^2$ coefficients, we get $a'(x)=-iG$. Setting $a(x) = \ri g(x)$, where
$g(x)$ is a real function, we obtain $G=-g'(x)$.
Inserting this $a(x)$ formula into the $\partial_x^1$ equation, we get
\[
b' = - g'g + \frac{\ri}{2}  g'' + n'.
\]
This equation can be integrated once, and we get
\[b = n -\frac{1}{2} g^2 + \frac{\ri}{2}  g' + c_1, \]
where $c_1$ is a constant.

Now we insert these $a(x)$ and $b(x)$ solutions into the $\partial_x^0$ equation. After simple algebra, this equation becomes
\[
n'g+2n g'=g^2 g'-\frac{1}{2} g'''-2 c_1 g'.
\]
Multiplying this equation by $g$, it can be rewritten as
\[
(ng^2)' = g^3g' -\frac{1}{2} g''' g- 2 c_1 g'g,
\]
from which we can solve the refractive index $n(x)$ as
\begin{equation}  \label{e:nII}
n = \frac{1}{4} g^2+ \frac{g'^2-2 g''g +c_2}{4g^2} - c_1,
\end{equation}
where $c_2$ is a real constant. The overall constant $c_1$ can be removed without loss of generality.

For the present second-order differential $\eta$ operator, the function $g(x)$ and the gain-loss profile $G(x)$ are also related by $G=-g'(x)$. Thus for any gain-loss profile $G(x)$, $g(x)$ is still given by Eq. (\ref{e:gxc0}), which contains a free real parameter $c_0$. Inserting this $g(x)$ into the refractive-index formula (\ref{e:nII}), we obtain the second family of complex potentials satisfying the similarity condition (\ref{Eq:LSimilar}) as
\begin{equation}  \label{Eq:PotentialTypeII}
V = \frac{1}{4} (\hat{g}+c_0)^2+ \frac{\hat{g}'^2-2\hspace{0.05cm} \hat{g}'' (\hat{g}+c_0) +c_2}{4(\hat{g}+c_0)^2} -iG,
\end{equation}
where $\hat{g}(x)$ is the cumulative gain-loss function defined below Eq. (\ref{e:gxc0}). For a given gain-loss profile $G(x)$, the refractive index in this potential has two free real parameters, $c_0$ and $c_2$, thus it has a wider range of tunability. This second family of potentials (\ref{Eq:PotentialTypeII}) has never been reported before to our best knowledge.

\begin{figure}[htb]
\includegraphics[width=8.3cm]{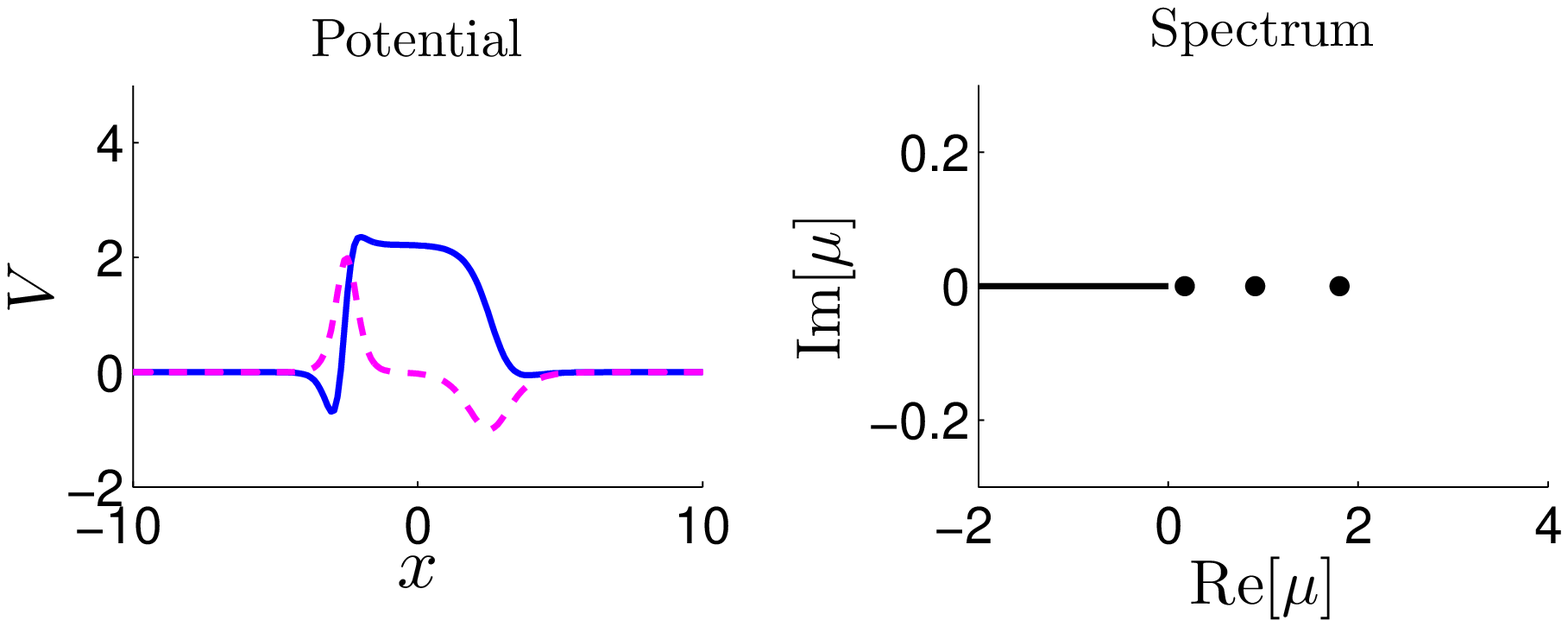}

\vspace{0.1cm}
\includegraphics[width=8.3cm]{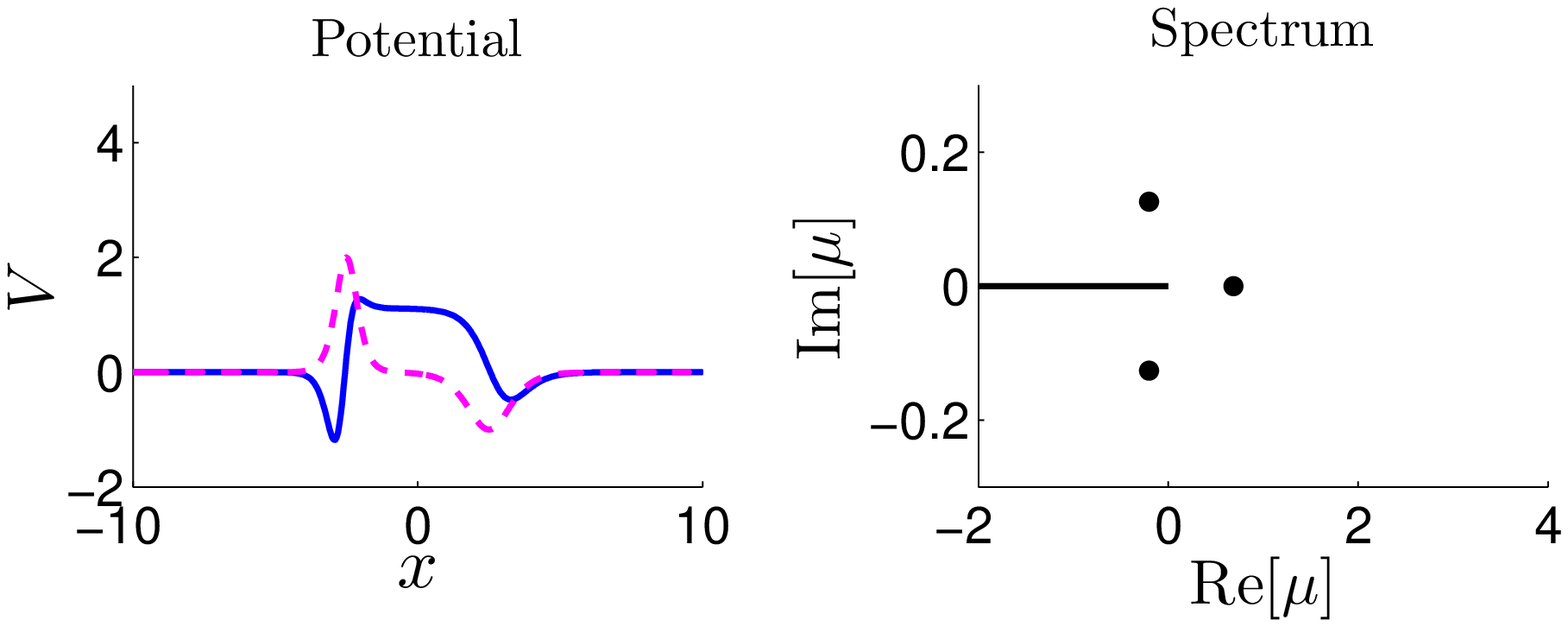}
\caption{Spectra of potentials (\ref{Eq:PotentialTypeII}) with gain-loss profile (\ref{Eq:GL}). Upper row: this potential with $c_0 = 1$ and $c_2=-1$; lower row: this potential with $c_0 = 1$ and $c_2=4$.  }
\label{Fig:SecondOrder}
\end{figure}

For an arbitrary gain-loss distribution $G(x)$, this second family of potentials also feature all-real spectra for a wide range of $c_0$ and $c_2$ values. For example, with the same asymmetric gain-loss distribution (\ref{Eq:GL}), the spectrum of this potential with $c_0=1$ and $c_2=-1$ is completely real, see Fig.~\ref{Fig:SecondOrder} (top panel). Fixing $c_0$ and decreasing $c_2$ will maintain the all-real spectra. If $c_2$ is increased above a certain threshold, phase transition occurs, where conjugate pairs of complex eigenvalues appear in the spectrum (see Fig.~\ref{Fig:SecondOrder}, bottom panel with $c_2=4$). For this bottom panel, an overall constant $(c_0^2+c_2/c_0^2)/4$ has been subtracted from the potential (\ref{Eq:PotentialTypeII}) so that the refractive index drops to zero at infinity.

\small
\normalsize

\begin{figure}[htb]
\includegraphics[width=8.3cm]{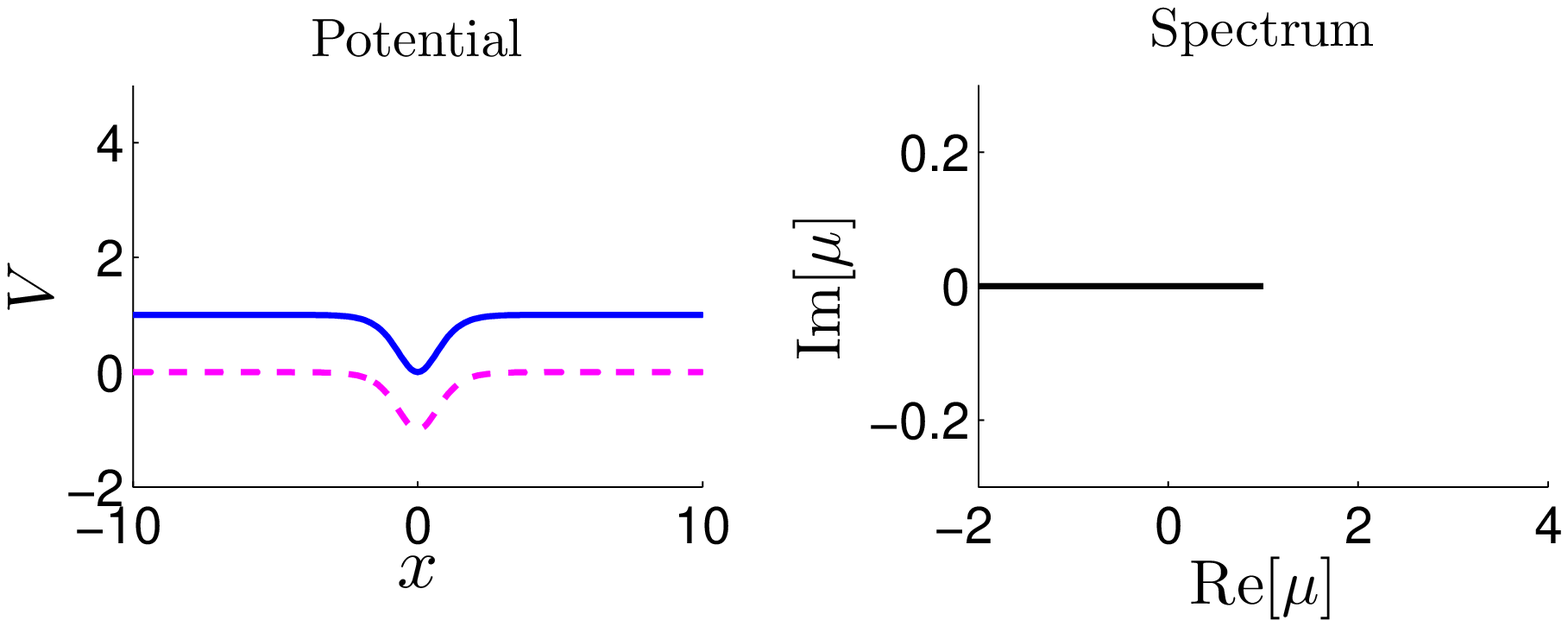}

\vspace{0.1cm}
\includegraphics[width=8.3cm]{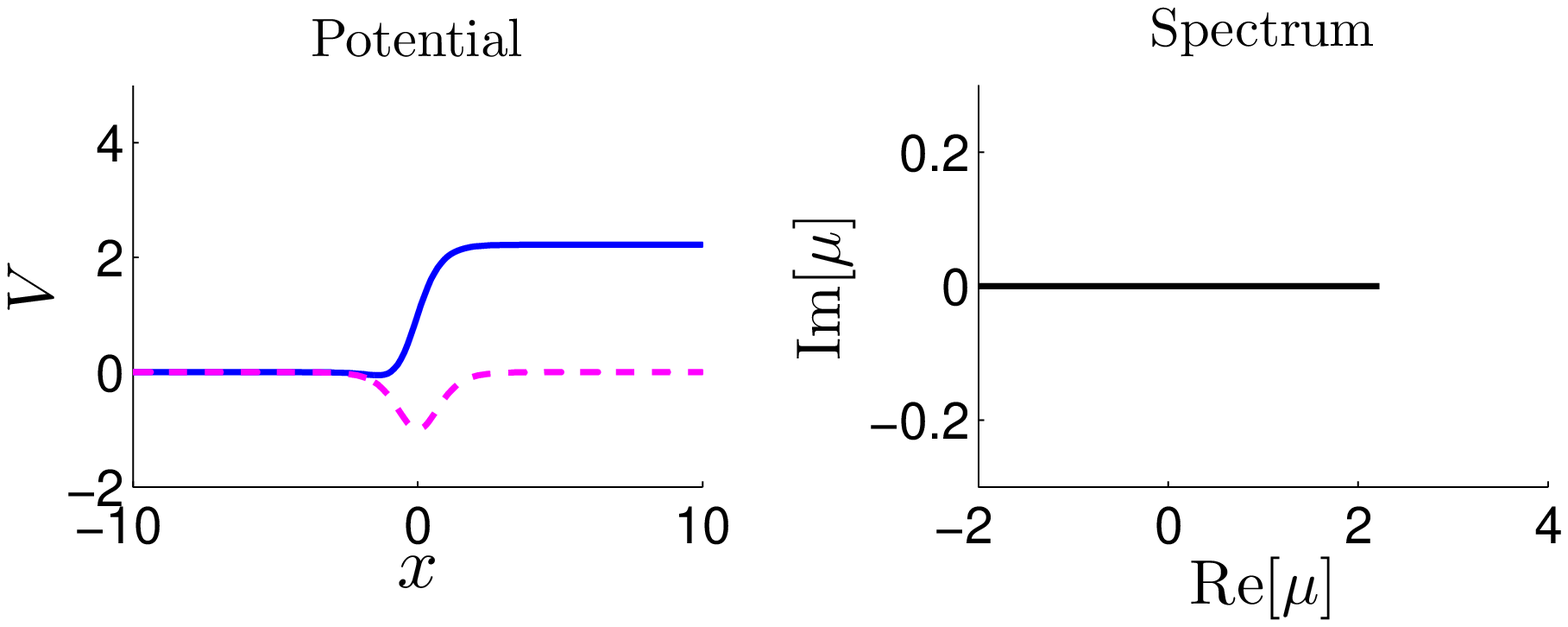}
\caption{Spectra of potentials with pure gain (\ref{Eq:GL2}). Upper row: potential (\ref{Eq:PotentialTypeI}) with $c_0 = 0$; lower row: potential (\ref{Eq:PotentialTypeII}) with $c_0 = 2$ and $c_2=-1$.  }
\label{Fig3}
\end{figure}

In the previous examples, all-real spectra were obtained for complex
potentials with both gain and loss. Interestingly, potentials
(\ref{Eq:PotentialTypeI}) and (\ref{Eq:PotentialTypeII}) can yield
completely real spectra even if there is only gain and no loss (or
vise versa). For example, if we take the pure-gain distribution
\begin{equation} \label{Eq:GL2}
G(x)  = \sech^2 \hspace{-0.02cm} x,
\end{equation}
then the spectra for the first type of potentials (\ref{Eq:PotentialTypeI}) with $c_0 = 0$ and the second-type of potentials (\ref{Eq:PotentialTypeII}) with $c_0 = 2$ and $c_2=-1$ are displayed in Fig.~\ref{Fig3}. It is seen that both spectra are completely real. This all-real spectra under pure gain is surprising, since it indicates that any initial wave cannot grow exponentially upon propagation even though the system has only gain and no loss. Our evolution simulation of the Schr\"odinger equation (\ref{e:SE}) with these complex potentials shows that an initial beam does not grow in amplitude at all. In the first example, with a dip in the index of refraction (see Fig. \ref{Fig3}, top panel), an initial Gaussian beam spreads out from the gain source in both directions so that light doesn't build up in any one place and the intensity remains bounded. In the second example, with a jump in the index of refraction, light is funneled away rightward (to the higher-index direction). Thus our results reveal that by properly engineering the refractive index distribution, beam amplification can be completely suppressed in systems with pure gain.

\small
\normalsize

In the above families of potentials, the differential $\eta$ operator was chosen to be the first-order or second-order. If $\eta$ is taken as the third- or higher-order operators, even more families of complex potentials with all-real spectra and phase transition could be constructed.

Our method of constructing non-\PT-symmetric complex potentials with
real spectra can be naturally extended to higher dimensions. Let us
consider paraxial light propagation in a three-dimensional (3D)
waveguide, which gives rise to a 2D Schr\"{o}dinger operator
\begin{equation} \label{e:V2D}
L=\partial_{xx}+\partial_{yy}+V(x,y).
\end{equation}
To construct complex potentials $V(x,y)$ with all-real spectra, we still impose the similarity condition (\ref{Eq:LSimilar}). As before, various choices of the differential operator $\eta$ will lead to different families of potentials. The simplest choice is a first-order $\eta$ operator
\begin{equation} \label{e:eta2D}
\eta=a(x,y)\partial_x+b(x,y)\partial_y+c(x,y).
\end{equation}
Substituting the above two equations into the similarity condition (\ref{Eq:LSimilar}), terms of third-order derivatives automatically balance. Balancing of second-order derivatives gives
\[
a_x=b_y=0, \quad a_y+b_x=0,
\]
which means that $a=a_0-\alpha y, b=b_0+\alpha x$, where $\alpha, a_0$ and $b_0$ are constants. For simplicity, we select $\alpha=0$. Then with a rotation of the $(x,y)$ coordinates and a scaling of $\eta$, the first-order $\eta$ operator (\ref{e:eta2D}) can be rewritten as $\eta=\partial_x+c(x,y)$. Finally, balancing of the remaining terms in (\ref{Eq:LSimilar}) leads to $c_x=-iG$, $c_y=0$, and $V_x=c_{xx}-2cc_x$. Setting $c=ig(x)$, where $g(x)$ is an arbitrary real function, then the 2D complex potential $V$ is found to be
\begin{equation}  \label{e:V2Dform}
V(x,y)=g^2(x)+f(y)+ig'(x),
\end{equation}
where $f(y)$ is another arbitrary real function. This 2D complex potential satisfies the similarity condition (\ref{Eq:LSimilar}), thus its spectrum can be completely real. Another way to see this all-real spectrum is that, since this potential is separable, its eigenvalues $\Lambda$ are of the form $\Lambda=\lambda_1+\lambda_2$, where $\lambda_1$ is an eigenvalue of the 1D operator $L_1=\partial_{xx}+g^2(x)+ig'(x)$, and $\lambda_2$ is an eigenvalue of another 1D operator $L_2=\partial_{yy}+h(y)$. We have known from the earlier text that the spectrum of $L_1$ can be all-real, and the spectrum of $L_2$ is naturally all-real since $L_2$ is Hermitian, thus the spectrum of the 2D potential (\ref{e:V2Dform}) can be all-real. Interestingly, this 2D potential coincides with the one reported in \cite{Yang_PRE2015}, where symmetry breaking of solitons was reported.

In the above calculations, if we let $\alpha\ne 0$ in functions $a$ and $b$, or we select higher-order differential $\eta$ operators, then further families of 2D complex potentials with completely real spectra can be derived.

In summary, we have outlined a method by which the spectrum in an optical potential can be made completely real for an arbitrary gain and loss distribution by judiciously constructing the index of refraction. This construction of the refractive index is not unique, since multiple families of refractive indices exist for the same gain and loss profile. Within these families, the spectrum can also undergo a phase transition.

While \PT-symmetric optics has garnered much interest recently in the optics community, the stringent \PT restriction on the gain and loss distribution limits its applicability. Our results show that the study of non-Hermitian optics with all-real spectra need not revolve around \PT-symmetry, since all-real spectra and phase transition can be realized for arbitrary gain and loss distributions (including optical media with pure gain and no loss). These findings open the door to a new research direction of non-\PT-symmetric optics. In addition, they imply that non-\PT-symmetric single-mode lasers with arbitrary gain and loss distributions in the laser cavity may be constructed \cite{PTlaser_Zhang, PTlaser_CREOL}. Beyond optics, our results also have interesting implications to non-Hermitian quantum mechanics, since wider classes of non-\PT-symmetric complex potentials become admissible now due to their all-real spectra. In addition, our results have ramifications to other physical areas such as Bose-Einstein condensates, where the mathematical model (the linear Gross-Pitaevskii equation) is the same as the Schr\"odinger equation (\ref{e:SE}) considered in this article \cite{Pitaevskii2003}.

This work was supported in part by the Air Force Office of
Scientific Research (Grant No. USAF 9550-12-1-0244) and the National
Science Foundation (Grant No. DMS-1311730).

\end{document}